\documentclass[a4paper]{jpconf}
\usepackage{graphicx}
\usepackage{amsmath}
\usepackage{xspace}
\usepackage{fancybox}
\usepackage{float}
\usepackage{subfloat}
\usepackage{enumitem}
\usepackage{color}
\usepackage{amsfonts}
\usepackage{array}
\usepackage[font=scriptsize]{caption}
\usepackage[subrefformat=parens,labelformat=parens]{subfig}
\usepackage[percent]{overpic}

\begin{document}

\title{Support Vector Machines and Generalisation in HEP}
\author{A. Bethani$^{1}$, A. J. Bevan$^{2}$, J. Hays$^{2}$ and T. J. Stevenson$^{2}$}
\address{$^{1}$LPSC, Grenoble, France. $^{2}$Queen Mary University of London, London, United Kingdom.}
\ead{a.j.bevan@qmul.ac.uk, t.j.stevenson@qmul.ac.uk}

\begin{abstract}
We review the concept of support vector machines (SVMs) and discuss examples of their use.
One of the benefits of SVM algorithms, compared with neural networks and decision trees is that they can be less susceptible to over fitting than those other algorithms are to over training. This issue is related to the generalisation of a multivariate algorithm (MVA); a problem that has often been overlooked in particle physics.
We discuss cross validation and how this can be used to improve the generalisation of a MVA in the context of High Energy Physics analyses. The examples presented use the Toolkit for Multivariate Analysis (TMVA) based on ROOT and describe our improvements to the SVM functionality and new tools introduced for cross validation within this framework.
\end{abstract}
\vspace{-1.0cm}

\section{Introduction}
Machine learning methods are used widely within High Energy Physics (HEP). One promising approach, used extensively outside of HEP for applications such as handwriting recognition, is that of Support Vector Machines (SVMs), a supervised learning model used with associated learning algorithms for multivariate analysis (MVA). Developed originally in the 1960s, with the current standard version proposed in 1995~\cite{vapnik}, SVMs aim to classify data points using a maximal margin hyperplane mapped from a linear classification problem to a possibly infinite dimensional hyperspace. However this means SVMs, like other MVA classifiers, have a number of free parameters which need to be tuned on a case by case basis. This motivates a number methods for ensuring the classifier is sufficiently generalised such that when used on an unseen dataset the performance can be accurately predicted.
In this paper a brief overview of SVMs is given in Section~\ref{sec:svm}, with an example using SVMs shown in Section~\ref{sec:checker}. Generalisation is discussed in Section~\ref{sec:gen} with an illustrative example of how this can improve performance given in Section~\ref{sec:check2}.

\section{Support Vector Machines}
\label{sec:svm}
\subsection{Hard Margin SVM}
\begin{figure}[htp]
  \centering
  \subfloat{\begin{overpic}[width=0.5\textwidth, trim=0 0 0 0,clip]{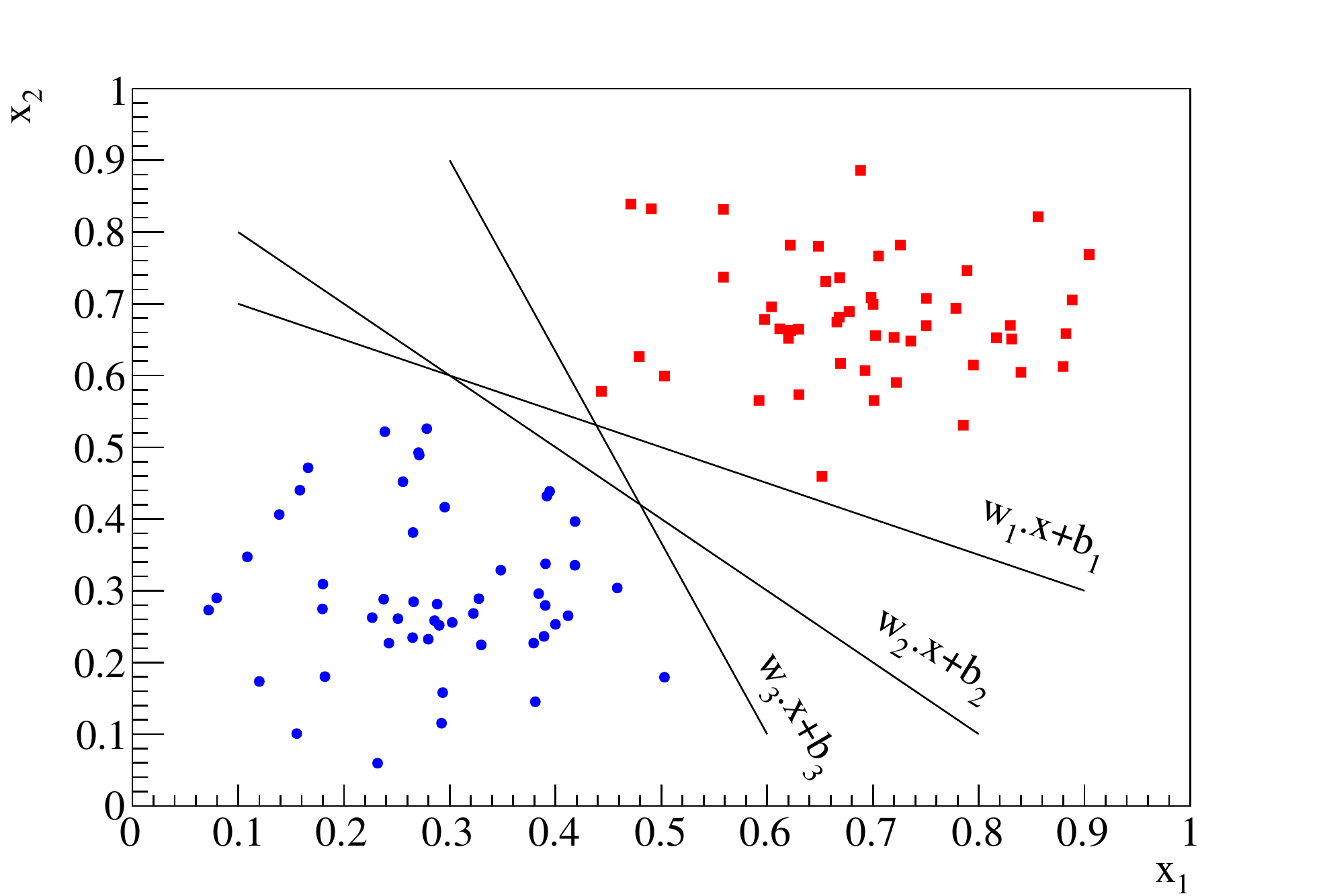}\put(0,0){(a)}\end{overpic}\label{fig:hardmargin}}
   \subfloat{\begin{overpic}[width=0.5\textwidth, trim=0 0 0 0,clip]{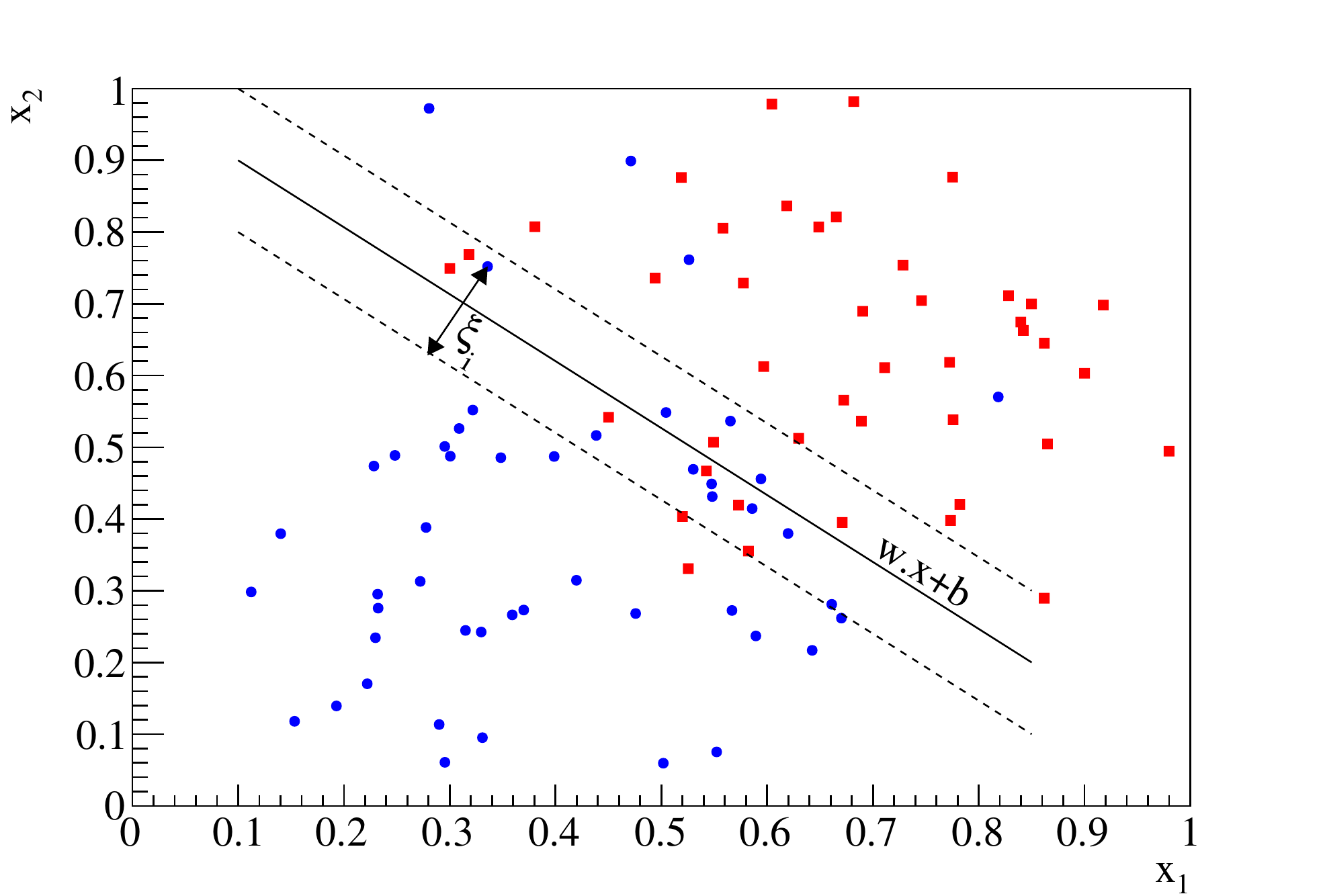}\put(0,0){(b)}\end{overpic}\label{fig:softmargin}}
  \caption{Illustrations of the (a) hard margin SVM where the solid line is the separating hyperplane, described by $\mathbf{w}$ and $b$,
  and the (b) soft margin SVM. 
  The slack parameters, $\xi_{i}$, allow for misclassification of points weighted by the tunable cost parameter, $C$.
  Dotted lines are the margins, with the SVM finding the maximum geometric margin, $\gamma$.
}
  \label{fig:results}
\end{figure}

Consider the problem of linear classification with the SVM where the training set, $\mathbf{x}$, is linearly separable. We define a separating hyperplane given by $\left<\mathbf{w}\cdot\mathbf{x}\right>+\,b = 0$, where $\mathbf{w}$, the weight vector, is perpendicular to the hyperplane, and $b$, the bias, determines the distance of the hyperplane from the origin (Fig.~\subref*{fig:hardmargin}). A decision function defined by $y_{i}=\mathrm{sign}(\left<\mathbf{w}\cdot\mathbf{x_i}\right>+\,b)$ is used to positively and negatively classify $\mathbf{x}_i$, the points in the training set. Without further constraint the choice of separating hyperplane is arbitrary and the geometric margin, $\gamma$, is introduced with the aim to maximise this. The functional margin, $\gamma_i$, for $\mathbf{x}_i$ is defined as
\begin{equation}
	\gamma_i = y_i(\left<\mathbf{w}\cdot\mathbf{x}_i\right>+\,b)\geq 1,
	\label{eq:functionalMargin}
\end{equation}
to impose the condition that no points lie within the margins. Considering two points that lie closest to the hyperplane referred to as support vectors (SVs), $\mathbf{x}^+$ and $\mathbf{x}^-$, with functional margins of $+1$ and $-1$ respectively, the geometric margin $\gamma$ can be defined as $\gamma = \frac{1}{2}\left(\left<\frac{\mathbf{w}}{\parallel\mathbf{w}\parallel}\cdot\mathbf{x^+}\right>-\left<\frac{\mathbf{w}}{\parallel\mathbf{w}\parallel}\cdot\mathbf{x^-}\right>\right) = \frac{1}{\parallel\mathbf{w}\parallel}$.
The problem is solved in dual space by minimising the Lagrangian
\begin{equation}
\widetilde{L}(\alpha)=\sum_{i=1}^{n}\alpha_{i} - \frac{1}{2}\sum_{i,j}\alpha_{i}\alpha_{j}y_{i}y_{j}\left\langle\mathbf{x}_{i}\cdot\mathbf{x}_{j}\right\rangle,
\label{eq:hardMarginLagrangian}
\end{equation}
for the parameters $\alpha_{i}$ (Lagrange multipliers) subject to $\alpha_{i} \geq 0$, where $\alpha_{i} $ is only non-zero for SVs, and $\sum_{i=1}^{n}\alpha_{i}y_{i} = 0$, providing the constraint equation. The full derivation leading to this minimisation and further details can be found in \cite{svmBook}.

\vspace{-1em}

\vspace{-1.8em}

\subsection{Soft Margin SVM}
The hard margin SVM is an important SVM algorithm, however it is not applicable in most real world cases. Issues arise if the data are not linearly separable due to being noisy, with data having small variations between different independent samples. This leads to the training becoming susceptible to over-fitting due to a few points in the training set, so some classification error must be allowed.
To achieve this the hard margin constraint is relaxed and points are allowed on the incorrect side of the decision boundary, with the misclassification described by two additional parameters, $\xi_{i}$ (slack) and $C$ (cost), where $\xi_i$ defines the distance from the correct margin to the $i$th incorrectly classified SV, and $C$ is a tunable weight parameter multiplying the sum of slack variables penalising the misclassified points (Fig.~\subref*{fig:softmargin}).
This reduces to the same dual space Lagrangian problem as for the hard margin SVM, Eq.~(\ref{eq:hardMarginLagrangian}), however the constraint on the Lagrange multipliers is now $0\leq\alpha_{i}\leq C$, while the constraint equation remains, $\sum_{i=1}^{n}\alpha_{i}y_{i} = 0$.

\subsection{Kernel Functions and Feature Spaces}
To this point the discussion of the SVM algorithm has been considered a linear decision boundary, however in most cases this is sub-optimal. In order to overcome this we can replace the inner product found in Eqns~(\ref{eq:functionalMargin}) and~(\ref{eq:hardMarginLagrangian}) with a kernel function (KF) $K(\mathbf{x},\mathbf{y})=\left\langle \phi(\mathbf{x})\cdot\phi(\mathbf{y})\right\rangle$ that maps the problem from the input space $X$ to a potentially higher dimensionality, implicit feature space $F=\{\phi(\mathbf{x})|\mathbf{x}\in X\}$, where the data may then be separable. 
In this process neither the feature map, $\phi(\mathbf{x})$, or the feature space, $F$, need to be known explicitly (known as the ``Kernel Trick"~\cite{svmBook}). This leads to specific properties required to define proper kernel functions in order for them to be used in this way.
The kernel function should satisfy $K(\mathbf{x},\mathbf{y}) = K(\mathbf{y},\mathbf{x})$
and $K(\mathbf{x},\mathbf{y})^2 \leq K(\mathbf{x},\mathbf{x})K(\mathbf{y},\mathbf{y})$. These are not strong enough conditions to guarantee the existence of F and so a KF is also required to meet Mercer's condition~\cite{Mercer}.\\

An example of a kernel meeting these requirements is the polynomial KF of Eq.~(\ref{eq:polynomial}) with two tunable parameters, the order $d$ and offset $c$. Another common choice of KF is the Radial Basis Function (RBF), Eq.~(\ref{eq:rbf}), where the distance between two SVs is computed and used as the input to a Gaussian with one tunable parameter $\Gamma=1/2\sigma^2$. The RBF does not take into account potential bandwidth differences between dimensions in the input data which may cause tension and the norm between SVs may result in loss of information. It can be extended to include a tunable parameter, $\Gamma_i$ for each dimension in the input space referred to as the multi-Gaussian KF, Eq.~(\ref{eq:multigauss}).  The KFs implemented in TMVA (along with products and sums thereof) are:

\vspace{-2em}
\begin{align}
K(\mathbf{x},\mathbf{y}) &= (\mathbf{x}\cdot\mathbf{y}+c)^{d} = \left(\sum_{i=1}^{\dim(X)}\mathbf{x}_{i}\mathbf{y}_{i}+c\right)^d,\label{eq:polynomial}\\
K(\mathbf{x},\mathbf{y}) &= e^{-\Gamma\parallel \mathbf{x} - \mathbf{y}\parallel^2},\label{eq:rbf}\\
K(\mathbf{x},\mathbf{y}) &= \prod_{i=1}^{\dim(X)} e^{-\Gamma_{i}(\mathbf{x}_i - \mathbf{y}_i)^2}.\label{eq:multigauss}
\end{align}
\vspace{-2.2em}

\subsection{Higgs Boson Example: $H\to \tau^+\tau^-$}
\label{sec:checker}
\begin{figure}[htp]
  \centering
   \subfloat{\begin{overpic}[width=0.5\textwidth, trim=0 0 0 30,clip]{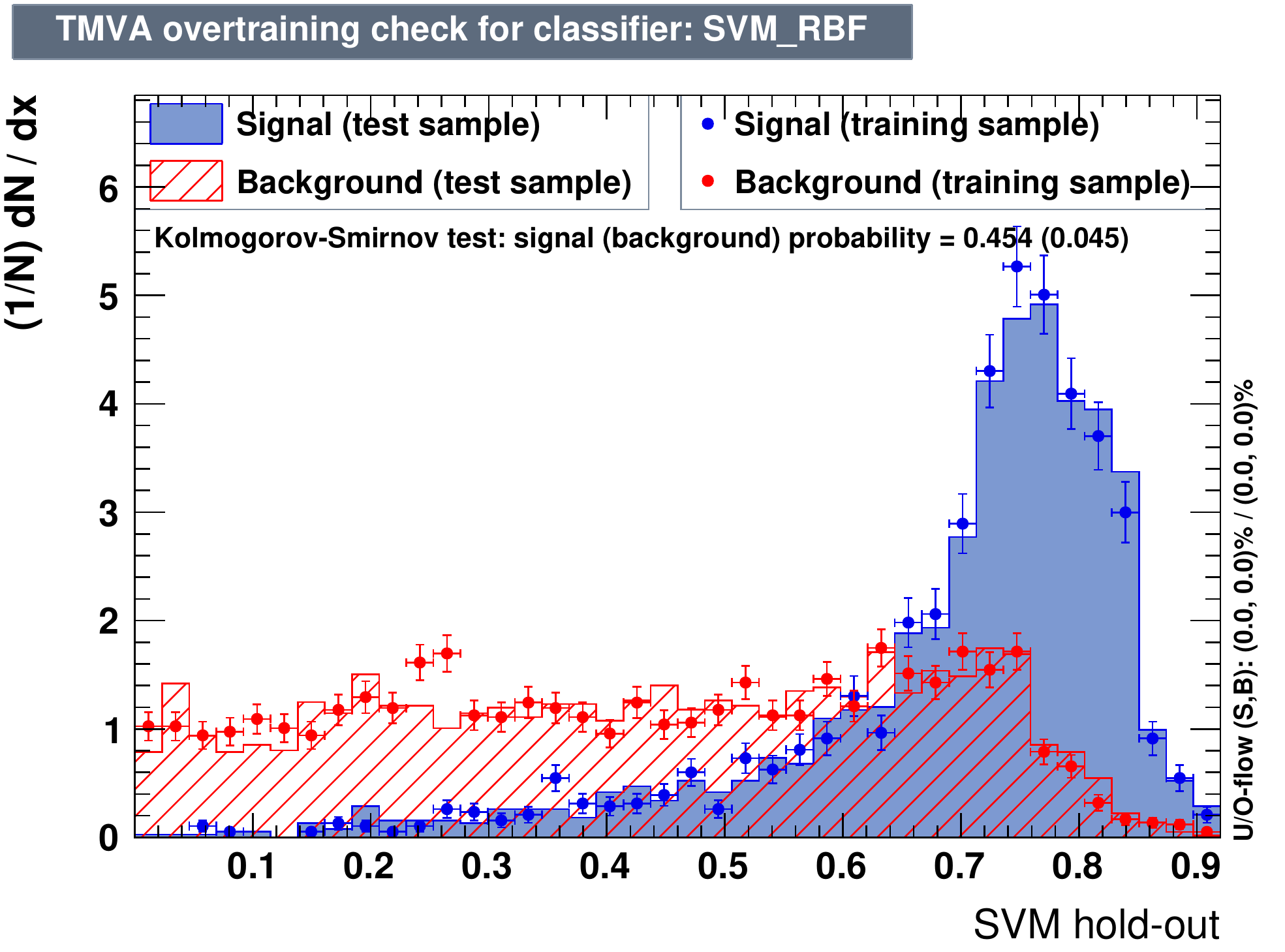}\put(0,0){(a)}\end{overpic}\label{fig:svmHoldout}}
   \subfloat{\begin{overpic}[width=0.54\textwidth, trim=0 0 0 30, clip]{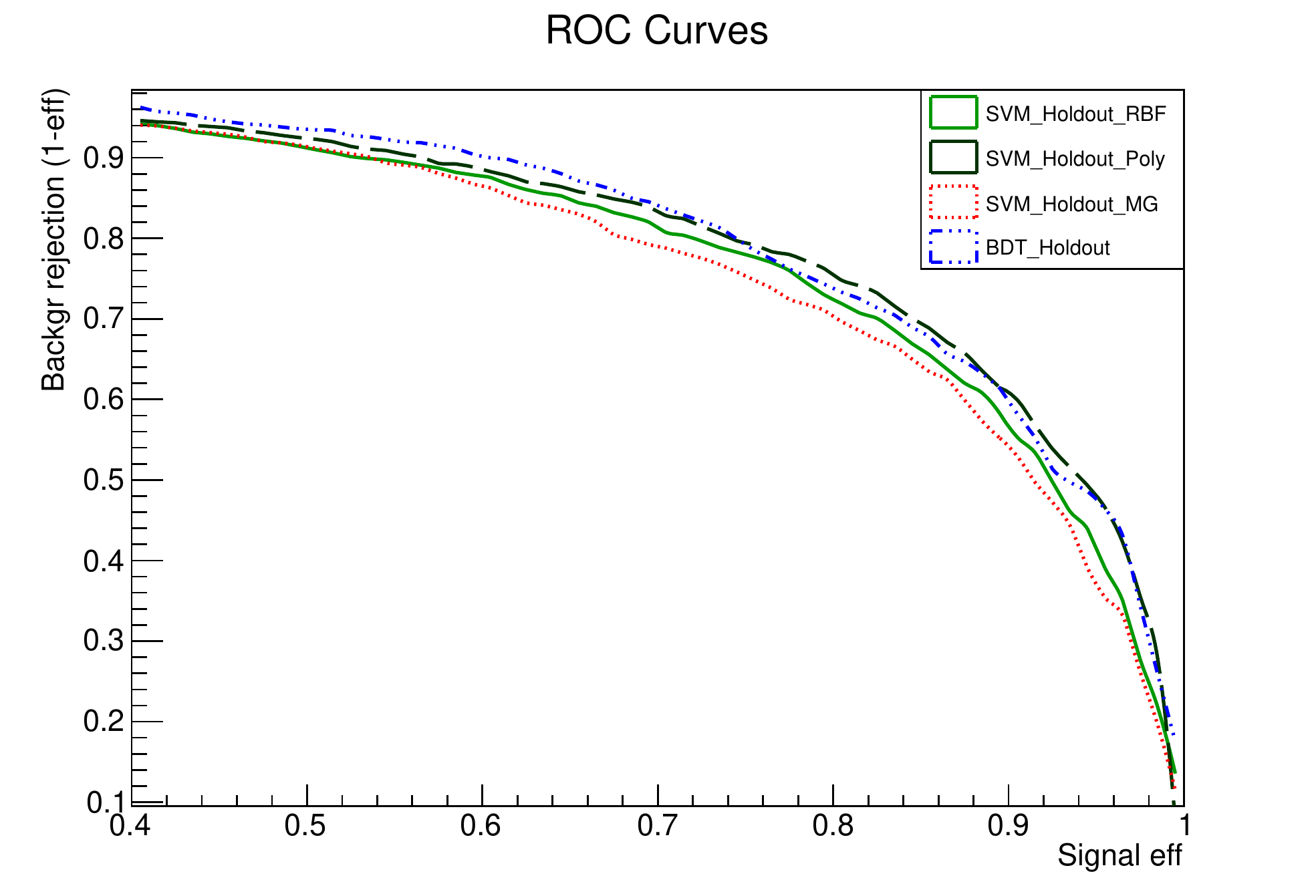}\put(0,0){(b)}\end{overpic}\label{fig:roc1}}
  \caption{(a) TMVA outputs for SVM with RBF kernel optimised, trained and tested on the Higgs Machine Learning Challenge dataset with hold-out validation method. (b) ROC curves for SVMs with polynomial, RBF and multi-Gaussian kernel functions and a BDT.}
  \label{fig:holdoutResults}
\end{figure}

The Higgs Machine Learning Challenge dataset~\cite{HiggsML}, with 6500 signal and 10000 background points is used to train SVMs with the KFs of Eqns~(\ref{eq:polynomial}-\ref{eq:multigauss}), as well as a Boosted Decision Tree (BDT)~\cite{BDT1}. The six variables offering the most discriminating power were used; $MMC$, $m_{T}(E_{T}^{miss},l)$, $\Delta R(\tau,l)$, $p_T^l/p_T^\tau$, $\phi$ Centrality and $E_{T}^{miss}$, see~\cite{HiggsML} for definition. The classifiers are optimised, trained and tested with TMVA~\cite{TMVA} using the hold-out validation method, discussed in Sec.~\ref{sec:holdout}. Considering the receiver operating characteristic (ROC) curves, signal efficiency versus background rejection shown in Fig.~\subref*{fig:roc1}, the classifiers all have similar performance for this problem. However it is not clear whether these solutions are fine tuned or optimal and so a further step is required to confirm if the results are generalised.

\vspace{-1em}

\vspace{-1.8em}

\section{Generalisation}
\label{sec:gen}
Due to the often non-transparent nature of multivariate techniques one requires confidence that the trained MVA is robust against over-fitting, such that when used on an unseen data sample the performance of the MVA is accurately predicted and therefore the MVA is generalised. In order to ensure this, validation techniques are required for both model selection, as most MVAs have at least one free hyper-parameter (HP), and for performance estimation, measured with a figure of merit such as the misclassification error. For an unlimited dataset these issues are trivial, as one could simply iterate through the HP space and models using a different subset of the data each time, evaluating the performance and choosing the model and configuration with the lowest error rate. In reality datasets are often much smaller than required.
Na\"{i}vely the entire dataset could be used to select and train the classifier and to estimate the error, however this leads to over-fitting or over-training as the classifier learns specific fluctuations in the dataset and subsequently performs worse on unseen data. This is more distinct in classifiers with a large number of tunable HPs. This also gives an overly optimistic estimation of the error.

\subsection{Hold-out Validation}
\label{sec:holdout}
A natural extension and potential way to overcome these issues is to split the dataset into two subsamples, one for training and one for testing. This is referred to as hold-out validation~\cite{CV}. The training sample is used to select optimal HPs, using a method such as back propagation for the Multilayer Perceptron~\cite{MLP}, and train the classifier. The testing sample is used to evaluate the performance. This method is not ideal as the performance differs depending on how the data is split, which may lead to misleading error estimates.

\subsection{k-fold Cross-validation}
\label{sec:kfold}
In circumstances where a relatively small dataset is available it may not be possible to reserve a large portion of data for testing and so the hold-out method may not be viable. Instead the approach of $k$-fold cross-validation (CV)~\cite{CV} can be used, performed as follows:

\begin{enumerate}
\item Split the data into $k$ equally sized, randomly sampled, independent datasets (or folds);
\item Train the classifier with the data from $k-1$ of the folds;
\item Test the classifier using the remaining fold;
\item Repeat steps (ii) and (iii) $k$ times for all permutations.
\end{enumerate}
The overall performance for the classifier is then given by the average error rate, $E=\frac{1}{k}\sum_{i=1}^{k}E_{i}$, where $E_{i}$ is the misclassification rate for each training. This method has the advantage of using the entire dataset for testing and training, offering potentially improved performance over the previous methods.

However it raises the question of the number of folds to be used. A large number of folds will give good estimation of the average error rate, as the bias of the estimator is small, but the variance is large and the computational time is large, whilst the reverse is true for a small number of folds. In reality the choice of number of folds should be motivated by the size of the dataset available for the problem, for example a sparse dataset may require the extreme of leave-$p$-out CV~\cite{CV} in order to train on as much data as possible.\\\\
Building on the $k$-fold method, ideally the data should be split into three statistically independent, randomly sampled datasets for training, testing and validation, with each set split into $k$-folds.
The purpose of the training set is to choose the model and to optimise the HPs for this model through iterative $k$-fold CV, with one set of HPs for each fold. These HPs are then tested via $k$-fold CV with the validation set to obtain their estimated performance. The training and validation samples are then combined and used to train the final classifier with the best performing HPs. The test sample, which has been reserved until this point, is then used to assess the performance of this final fully trained classifier.\\\\
Taking the best performing classifier from the optimisation process may not be the optimal solution or give the desired output, due to potential pathologies in the distributions. This also involves discarding a large number of trained classifiers from the process, so instead of using all HPs sets from the optimisation step, $k$ classifiers can be trained and their performance estimated. These can then be used to give an averaged output on the testing sample which can help reduce fluctuations in the final distributions, improving agreement between the performance on the training and testing samples.

\subsection{Higgs Boson Example: $H\to \tau^+\tau^-$}
\label{sec:check2}
\begin{figure}[htp]
  \centering
   \subfloat{\begin{overpic}[width=0.5\textwidth, trim=0 0 0 30,clip]{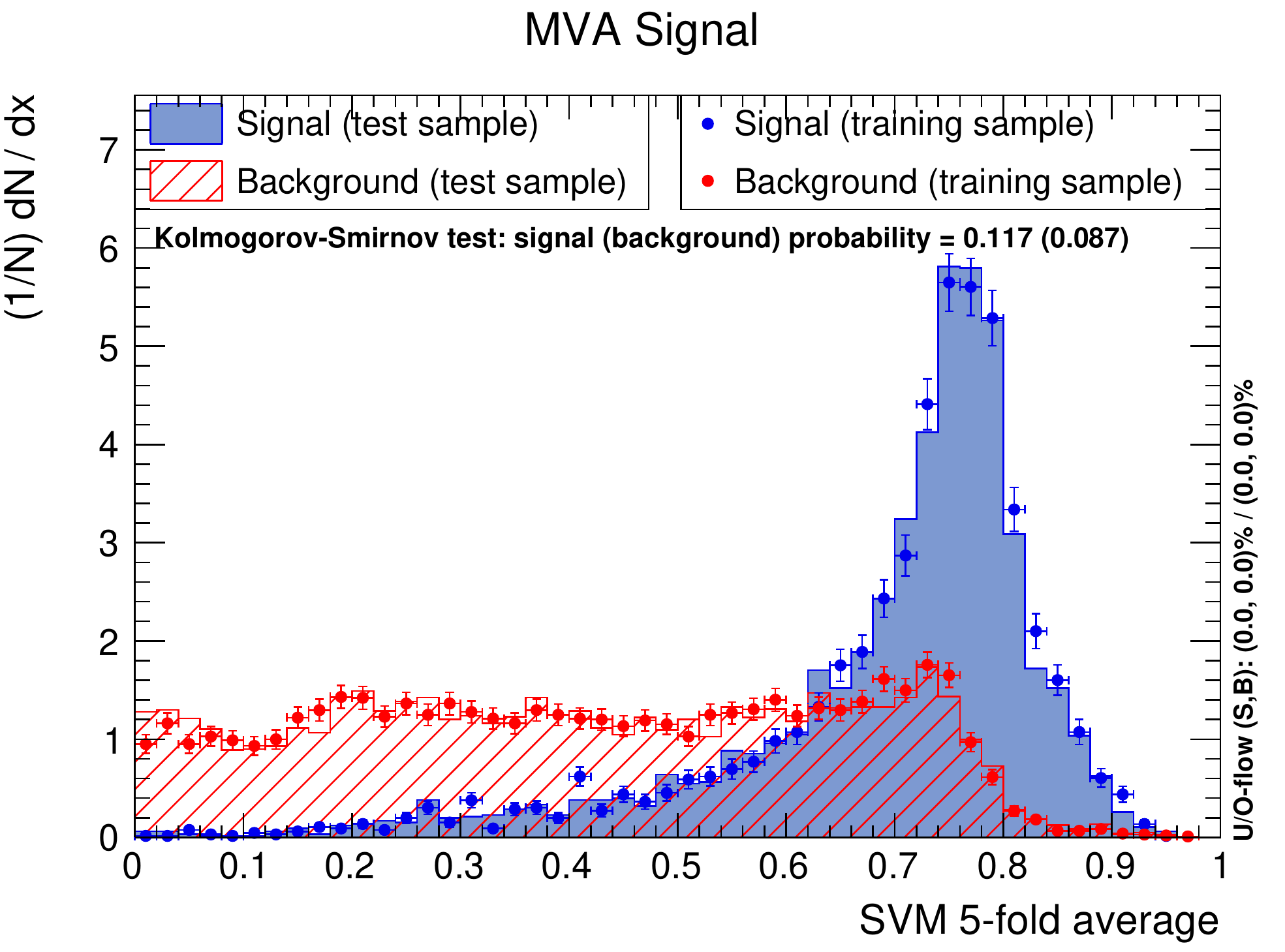}\put(0,0){(a)}\end{overpic}\label{fig:svmAv}}
   \subfloat{\begin{overpic}[width=0.54\textwidth, trim=0 0 0 30, clip]{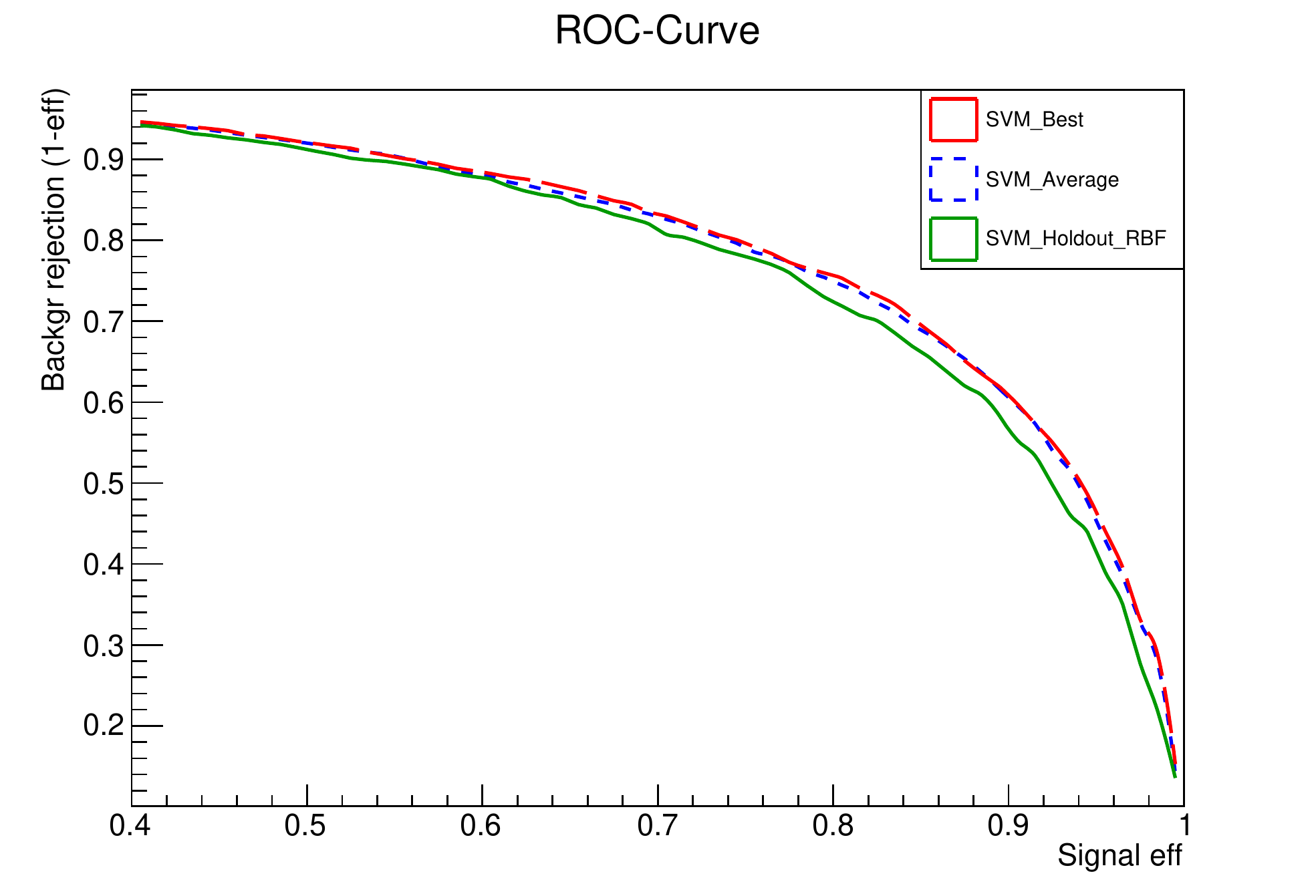}\put(0,0){(b)}\end{overpic}\label{fig:finalRoc}}
  \caption{(a) TMVA output for a SVM with RBF kernel optimised, trained and tested on the Higgs Machine Learning Challenge dataset using $5$-fold validation method averaging the classifiers from each fold. (b) ROC curves showing improved performance of the best and averaged $5$-fold classifiers over the hold-out method classifier.}
  \label{fig:results}
\end{figure}

Using the Higgs Machine Learning Challenge dataset, discussed in Sec.~\ref{sec:checker}, SVMs with RBF KFs were optimised, trained and tested using $5$-fold CV to obtain the best performing and averaged SVMs, as well as using the hold-out validation method for comparison. The ROC curve comparing the different methods, Fig.~\subref*{fig:finalRoc}, shows improvement for the $k$-fold CV classifiers over the hold-out method, both in the performance and shape of the curves. It is also worth noting that for the best performing and averaged classifiers from the $k$-fold CV, Fig.~\subref*{fig:svmAv}, show improved agreement between the training and test samples than for the hold-out validated method, Fig.~\subref*{fig:svmHoldout}, illustrating how this can lead to a more generalised result.

\vspace{-1em}

\vspace{-2.5em}

\section{Summary}
A brief overview of Support Vector Machines was presented, with an example showing similar performance to that of a BDT. However it is not clear without further checks as to whether the MVAs are sufficiently generalised. Hence a multistage cross-validation procedure has been outlined, which for the same example shows better performance as well as better agreement between the training and testing samples in the output distributions. This functionality is now available in the github version of ROOT.


\vspace{-1em}

\section*{References}
\begin{thebibliography}{9}
\bibitem{vapnik} C. Cortes and V. Vapnik, Machine Learning Volume \textbf{20} Issue 3 (1995).
\bibitem{svmBook}J. Shawe-Taylor and N. Cristianini, Support Vector Machines and other kernel-based learning methods, Cambridge University Press (2000).
\bibitem{Mercer} J. Mercer, Philosophical Transactions of the Royal Society A \textbf{209} Issue 441$-$458 (1909).
\bibitem{HiggsML}
C.~Adam-Bourdarios, G.~Cowan, C.~Germain-Renaud, I.~Guyon, B.~Kégl and D.~Rousseau,
``The Higgs Machine Learning Challenge,'' J.\ Phys.\ Conf.\ Ser.\  {\bf 664} (2015) No.7,  072015.
See also URL:
http://higgsml.lal.in2p3.fr/documentation/ (2014).
\bibitem{BDT1}L. Breiman, J. Friedman, R. Olshen, and C. Stone, Classification and Regression Trees, Wadsworth International Group (1984).
\bibitem{TMVA}A. Hoecker et al., PoS ACAT 040 (2007) 040.
\bibitem{CV}S. Arlot and A. Celisse, Statistics Surveys Vol. \textbf{4} 4079 (2010).
\bibitem{MLP}R. Rojas, Neural Networks: A Systematic Introduction, Springer Science $\&$ Business Media (1996).
\end{thebibliography}
\end{document}